\documentstyle[aps,prc,preprint,epsfig]{revtex}   

\begin{document}

\preprint{KUNS-1748,TMUP-HEL-0018,DUKE-TH-01-206}

\draft
\title{Experimental Signatures of Anomaly Induced DCC Formation}
\author{
Masayuki Asakawa\\
{\it Department of Physics, Kyoto University,
Kyoto, 606-8502, Japan} \\
Hisakazu Minakata\\
{\it Department of Physics, Tokyo Metropolitan University \\
Hachioji, Tokyo 192-0397, Japan}\\
Berndt M\"uller\\
{\it Department of Physics, Duke University,
Durham, NC 27708-0305, U.S.A.}\\
}

\maketitle

\begin{abstract}
We discuss characteristic experimental signatures related to the
formation of domains of disoriented chiral condensate (DCC)
triggered by the axial anomaly in relativistic heavy ion collisions.
We predict that the enhancement of the fraction of neutral pions 
compared to all pions depends on the angle of emission with respect 
to the scattering plane and is concentrated at small transverse 
momentum and small rapidity in the center-of-mass frame.
The anisotropy with respect to the reaction plane is also
observable in the inclusive photon distribution.
\end{abstract}
\medskip



The experimental search for domains of disoriented chiral
condensate (DCC) \cite{DCC} in high-energy heavy-ion collisions 
is under way. The first results were negative \cite{WA98,Minimax},
but this may be due to a variety of experimental limitations,
such as beam energy, pion momentum range, and event selection.
In two previous papers, we demonstrated that the axial anomaly
may enhance the formation of DCC domains in non-central nuclear
collisions \cite{MM96,AMM98}. Here we present an important new
result concerning the observational characteristics of DCC's created
by the anomaly mechanism. We show that they will exhibit a very
specific property, in that the ratio of neutral to charged pions
will depend on the angle of emission with respect to the nuclear
scattering plane. 

The anomaly mechanism for the creation of DCC domains is based
on the effective interaction
${\cal L}_{\rm int} = (\alpha/\pi f_\pi)\, {\vec E}\cdot{\vec B}\, 
\phi^0$ .
In non-central relativistic heavy ion collisions this interaction
generates a (classical) source for the neutral pion field $\pi^0$ 
in the form of an almost instantaneous ``kick''.
The short duration of the anomalous interaction, of the order of 
collision time, begs the question whether the use of the 
effective pion field is justified, or whether the electromagnetic 
field should rather be microscopically coupled to the quark 
degrees of freedom.  In response, we recall that the 
anomaly term  precisely describes the low-energy 
limit of the interaction between on-shell photons and the quarks 
in the neutral pseudoscalar channel. The short time scale of 
the anomalous interaction is effectively taken into account by 
using an uniformly ``kicked'' initial field configuration in our 
previous treatment in refs. \cite{MM96,AMM98}.

For far off-shell photons, the coupling is suppressed by a form 
factor \cite{FFpi}, but this effect is expected to be small here, 
because the coherent nuclear electromagnetic field corresponds to 
nearly on-shell virtual photons. Furthermore, we only consider the 
effect of the anomalous interaction on pion field modes with momenta 
far below the chiral symmetry-breaking scale 
$\Lambda_{\rm ch}\approx 1$ GeV.
Finally, the temperature dependence of the anomalous electromagnetic 
interaction of the neutral pion \cite{Pi96} needs to be considered. 
Strictly speaking, the anomaly provides a temperature-independent 
source term for the dynamics of the axial current $j_5^\mu$. As 
the temperature approaches the critical temperature $T_c$ for chiral 
symmetry breaking, the axial current is less and less dominated by 
single-pion excitations. Accordingly, the pion coupling to two photons
decreases and vanishes at $T_c$ \cite{PT97}. In principle, the
treatment of the anomaly term should be based on the full 
dynamics of the axial current, rather than the pion field
alone. However, since we are interested in the effects of the
interaction at very early times (the ``anomaly kick'' occurs
before thermalization) and at late times (after the assumed
rapid quench to conditions below the QCD critical temperature),
where the axial current is again dominated by the dynamics of
the pion field, the treatment in terms of the low-energy
effective interaction  appears to be a plausible
first approximation.

We now turn to our original question of the possible characteristic
signatures of the DCC formation triggered by the anomaly effect.
We here focus on the (previously ignored) correlation between the 
ion scattering plane and the $\pi^0/\pi$ ratio.
Due to the linear dependence of the electromagnetic invariant
${\vec E}\!\cdot\!{\vec B}$ on the impact parameter, the chiral anomaly
generates a ``kick'' on the collective pion field only in collisions
with a nonzero impact parameter \cite{MM96,AMM98}.
In these non-central collisions
one can use collective flow patterns of the produced hadrons to 
define a collision plane on an event-by-event basis \cite{flow}.
We choose the $x$-$z$ plane as the collision plane, taking the 
$z$-direction along the beam axis, and the $y$-axis normal to the 
collision plane. Because ${\vec E}\!\cdot\!{\vec B}$ is an odd 
function of $y$, the anomaly term has a characteristic dependence 
on the momentum component normal to the collision plane, $k_y$, 
rather than on the momentum components $k_x$ and $k_z$.

Therefore, one would expect that the coherent pion field excitation
due to the anomaly kick would result in the preferential emission 
of pions with values of $k_y$ that are small (reflecting the spatial 
coherence of the source) but nonzero. In fact, there should be no 
effect of the kick at all to pions with $k_y=0$,
since $\int_{-\infty}^{\infty} dy {\vec E}\!\cdot\!{\vec B} =0$
for a fixed $x$ and $z$. This argument remains valid also 
beyond perturbation theory because of the coherence of the source.
Following the procedure outlined in Section IV. E of
ref.~\cite{AMM98}, we have evaluated the $\pi^0/\pi$ ratio
as a function of the azimuthal angle $\phi$ around the beam axis. 
The direction $\phi=0^{\circ}$ coincides with our choice of the 
$x$-axis. According to the above, we expect an increase in the
$\pi^0/\pi$ ratio for emission near the angles $\phi=90^{\circ}$
and $\phi=270^{\circ}$.

\begin{figure}
\centerline{\epsfig{file=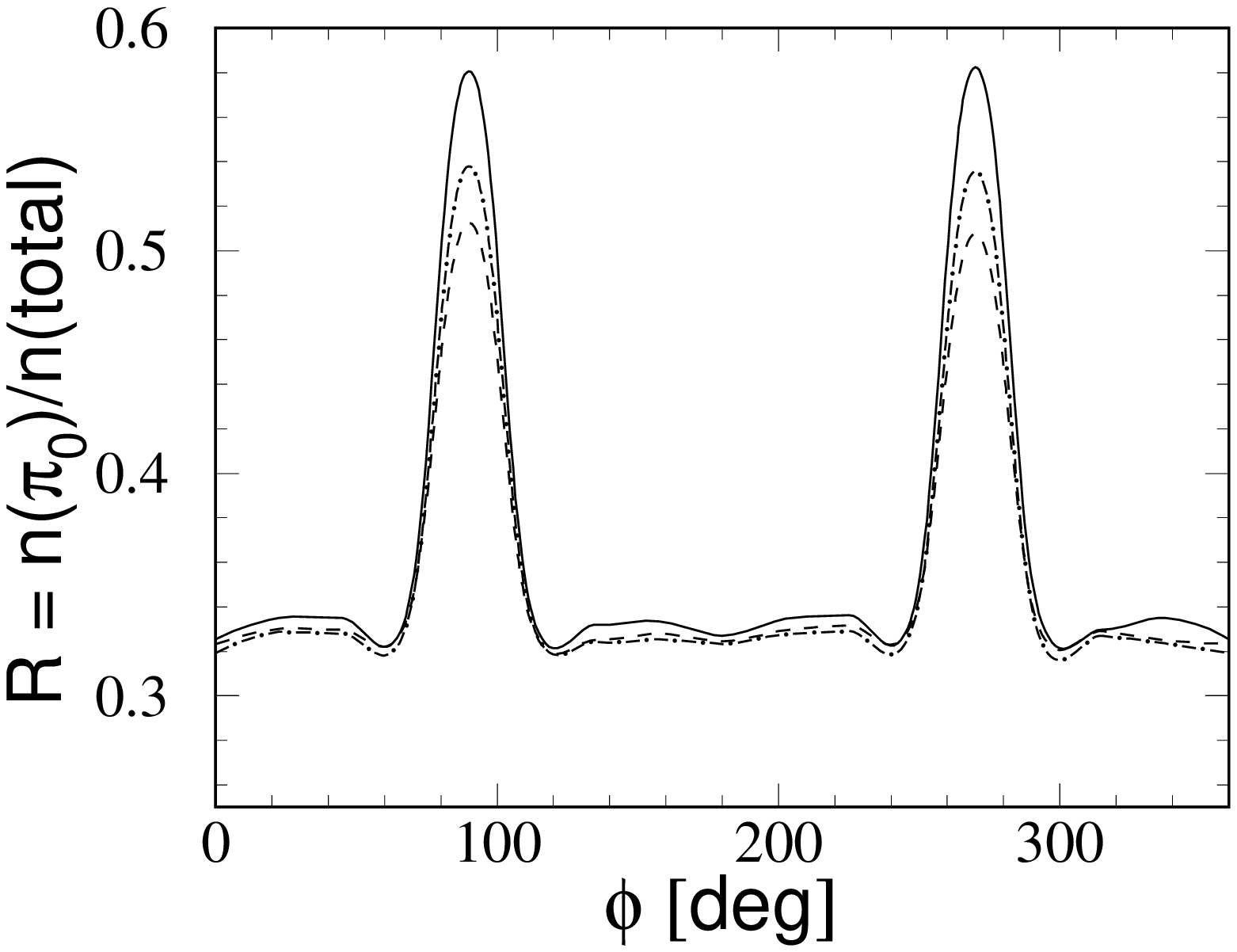,width=0.9\linewidth}}
\vskip 1cm
\caption{
  Azimuthal angle $\phi$-dependence of the ratio
  $n_{\pi^0}/n_{\pi}$ for $R_0 =\infty$ at three different times.
  Solid line: $\tau=1$ fm;
  dashed line: $\tau=6$ fm;
  dash-dotted line: $\tau=11$ fm.
  The two maxima lie in the direction normal to the scattering
  plane and are an expression of the spatial variation of the
  axial anomaly source term.}
\label{r=inf}
\end{figure}

\begin{figure}
\centerline{\epsfig{file=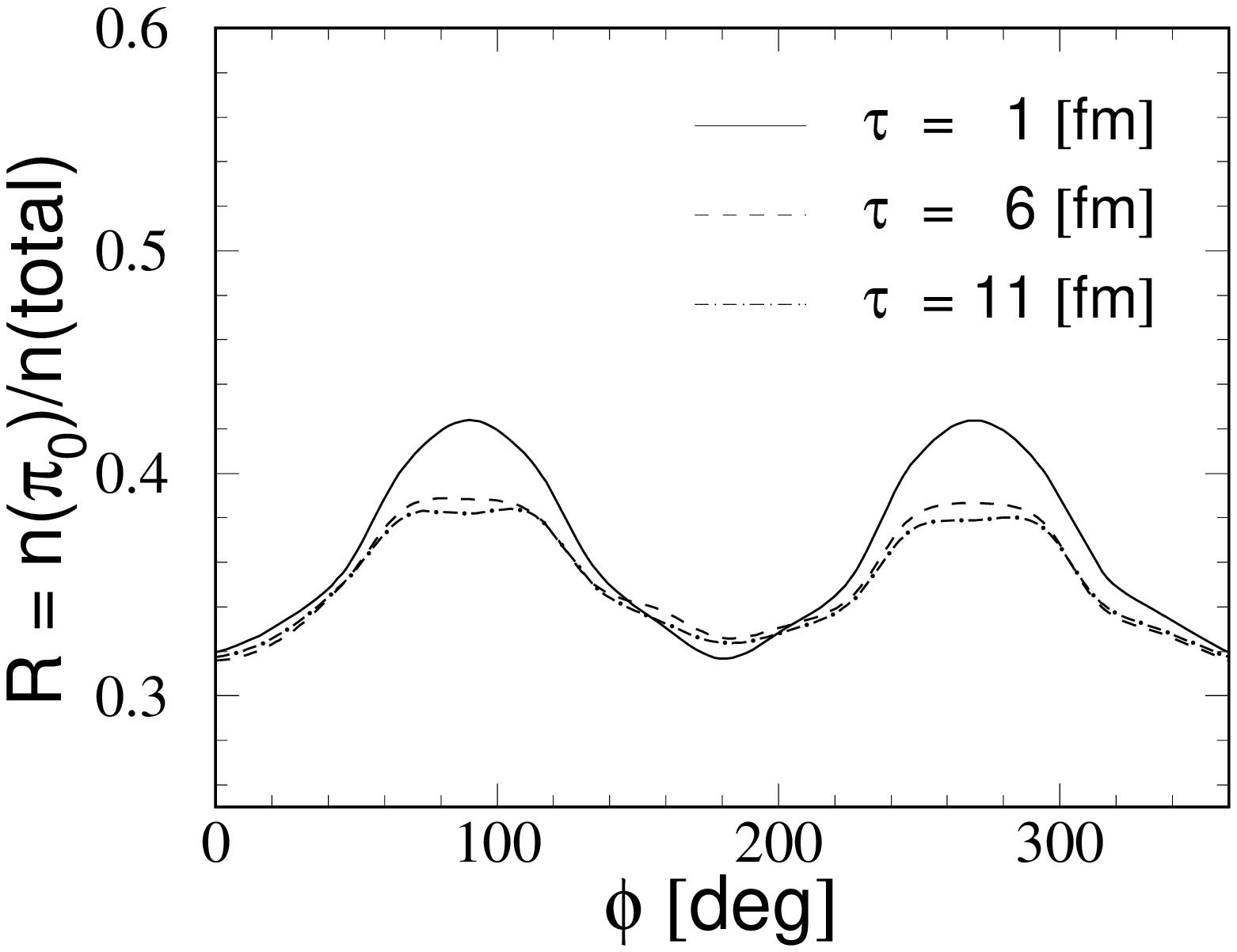,width=0.9\linewidth}}
\vskip 1cm
\caption{
  Azimuthal angle $\phi$-dependence of the ratio
  $n_{\pi^0}/n_{\pi}$ for $R_0 =5$ fm at three different times.
  Solid line: $\tau=1$ fm;
  dashed line: $\tau=6$ fm;
  dash-dotted line: $\tau=11$ fm.
  The two maxima lie in the direction normal to the scattering
  plane and are an expression of the spatial variation of the
  axial anomaly source term.}
\label{r=5}
\end{figure}

In order to test this hypothesis, we have generated 1000 events
following the procedure described in ref.~\cite{AMM98}. 
The anomaly kick parameter $a_n$ defined in ref.~\cite{AMM98}
has been taken to be 0.1. The size of the anomaly kick expected
in Au + Au collisions at $\sqrt{s} = 200$ GeV/A
at the Relativistic Heavy Ion Collider (RHIC) at Brookhaven
National Laboratory is of this order.
We did calculations both for an infinitely extended system
with $R_0 = \infty$ and for a finite system
with $R_0 = 5$ fm in the transverse direction, where
$R_0$ is defined in eq.~(22) of ref.~\cite{AMM98} and represents
the transverse extension of the system.
The other parameters are the same as those used in ref.~\cite{AMM98}.
Our results are shown in Figs.~\ref{r=inf} and \ref{r=5},
presenting the dependence of
the ratio $n_{\pi^0}/n_{\pi}$ of low momentum pions
($|{\vec k}|<250$ MeV) as a function of the azimuthal angle
$\phi$. The three different curves represent the results
at different times: $\tau=1$ fm (initial time, solid line),
$\tau=6$ fm (dashed line), and $\tau=11$ fm (dash-dotted line).

The angular distributions clearly show the expected behavior.
As described in our earlier work, the $\pi^0/\pi$ ratio is
overall enhanced for the low momentum pions (see Fig. 20 of
ref. \cite{AMM98}), but the enhancement exhibits strong peaks
near $\phi = 90^{\circ}$ and $\phi = 270^{\circ}$, i.e.~for emission normal
to the scattering plane. The directional peak structure
somewhat diminishes with time, but clearly survives until the
final moments in our calculations. The magnitude of the effect
and the width of the peaks depend on the size of the region 
affected by the coherent excitation, indicating that
the spatially coherent source term provided by the axial
anomaly is an essential ingredient of the effect.

The predicted azimuthal anisotropy of the $\pi^0/\pi$ ratio
provides a clear, and most likely unique, signature of DCC
domain formation triggered by the chiral anomaly. Its detection
relies on the ability to determine the orientation of the
collision plane by means of collective flow patterns in the
distribution of emitted particles. At high energies, the flow 
in the central rapidity region
exhibits an elliptical (or quadruple) distribution in the 
azimuthal angle $\phi$ \cite{olli,flow}. This effect
has been observed in Pb+Pb collisions at the CERN-SPS \cite{sps} 
and has recently also been confirmed for Au+Au collisions at 
RHIC \cite{star}. Both elliptical 
flow and the anomaly effect are characteristic for collisions 
with a nonzero impact parameter and absent in central collisions.

\begin{figure}
\centerline{\epsfig{file=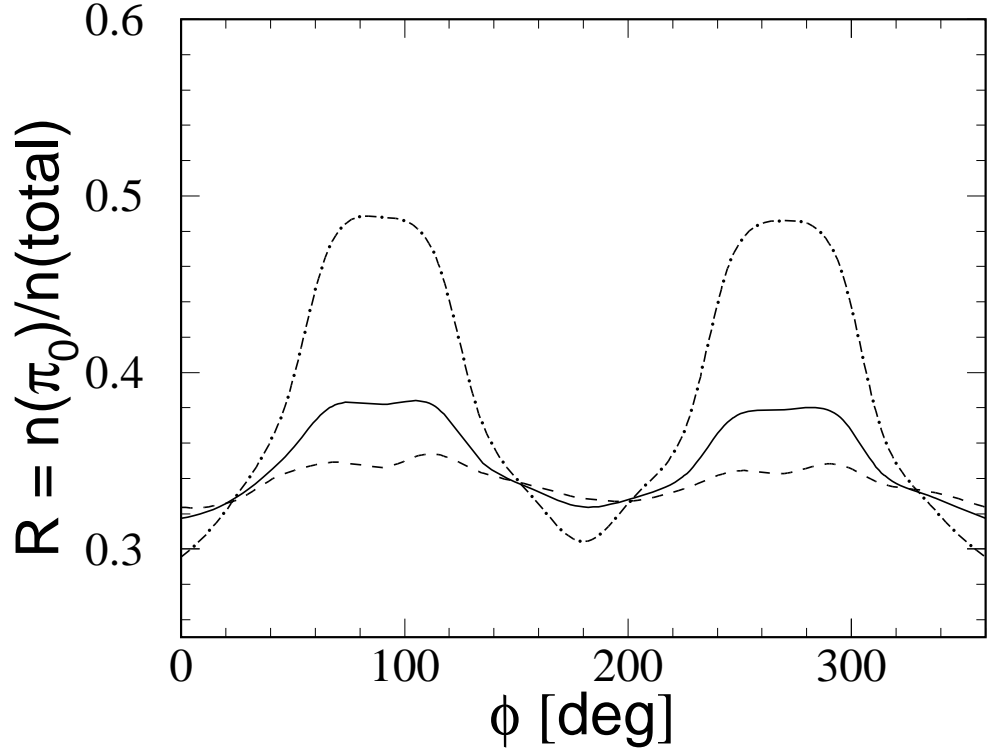,width=0.9\linewidth}}
\vskip 1cm
\caption{
  Dependence of the $\pi^0/\pi$ enhancement on the magnitude
  $a_{n}$ of the anomaly kick. $\tau$ is fixed at 11 fm.
  Dashed line: $a_n = 0.05$;
  solid line: $a_n = 0.1$;
  dash-dotted line: $a_n = 0.2$ }
\label{Zdep}
\end{figure}

Another characteristic signature for anomaly induced DCC formation 
would be the strong $Z$-dependence of the low-$k_y$ enhancement, 
which should serve as a confirmation of the electromagnetic nature 
of the anomaly effect. To reveal this dependence, we repeated the 
simulation with $R = 5$ fm for three different values of $Z$, 
expressed as three different values of the kick amplitude $a_{n}$. 
The results are shown in Fig.~\ref{Zdep}, presenting the dependence 
of the ratio $R \equiv n_{\pi^0}/n_{\pi}$ of low momentum pions
($|{\vec k}|<250$ MeV) as a function of the azimuthal angle
$\phi$. $\tau$ is fixed at 11 fm. The three different curves of 
$R$ represent the results at different $a_n$: 0.05 (dashed line), 
0.1 (solid line), and 0.2 (dash-dotted line).

If we fit the peak value of $R$ by using the function 
$c + b (a_n)^{\alpha}$ we obtain $\alpha \simeq 1.8$. 
Since  $a_{n}$ scales as $Z^{2}$, our simulation indicates that 
the $Z$-dependence of the anomaly effect is roughly consistent 
with $Z^4$ dependence. Assuming that the directional dependence
of the neutral-to-charged pion ratio is observed, it will be 
interesting to study the $Z$-dependence of this effect. Here
we note only that, in practice, the $Z$-dependence will be
partially obscured by the unavoidable dependences on various 
other experimental parameters, e.g. the nuclear radii, the 
impact parameter, and the initial energy density, which all
vary with nuclear size. These factors must be taken into 
account in the interpretation of the experimental data.

\begin{figure}
\centerline{\epsfig{file=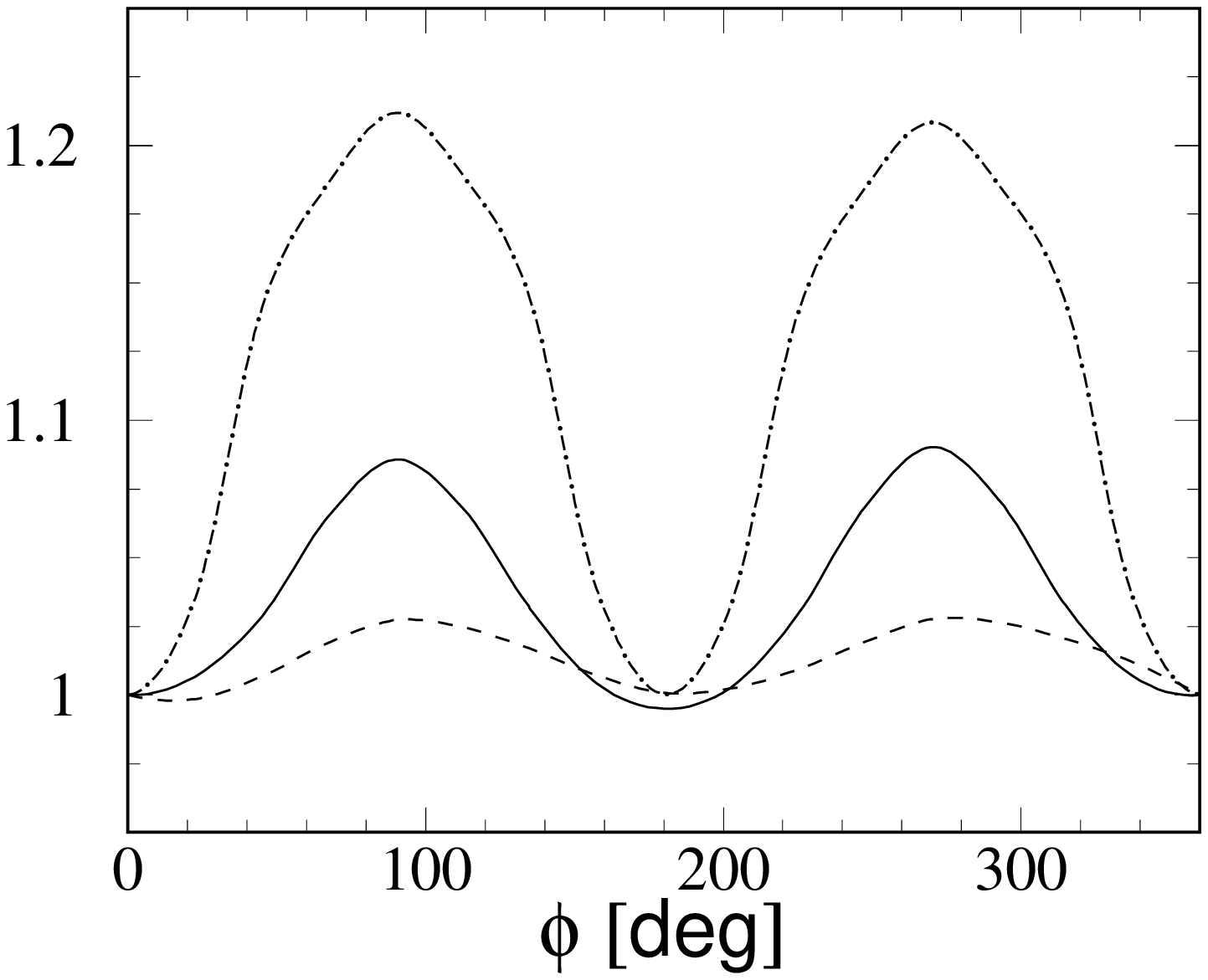,width=0.9\linewidth}}
\vskip 1cm
\caption{
  Dependence of azimuthal distribution of inclusive photons 
  from $\pi^0$-decay on the magnitude $a_{n}$ of the anomaly 
  kick, for all photons in the range $0 < k_T < 500$ MeV.
  The parameters for the various lines are the same as in
  Fig.~\protect{\ref{Zdep}}. The distribution is normalized
  to unity at $\phi=0$.}
\label{gammas}
\end{figure}

An important phenomenological question is whether the anisotropy
of the neutral pion distribution gives rise to an anisotropy of
the inclusive photon distribution. If this were the case, its
experimental observation would be greatly simplified, as it
would not be necessary to reconstruct the $\pi^0$ distribution
from two-photon coincidences. In the high-multiplicity
environment of relativistic heavy-ion collisions, such a 
reconstruction is only possible on a statistical basis. We
have therefore, calculated the azimuthal dependence of the
inclusive photon distribution resulting from the calculated
neutral pion distribution. The result, shown in Fig.~\ref{gammas}
for photons with $0 < k_T < 500$ MeV, clearly shows that the 
azimuthal anisotropy persists in the photon yield. Most of the
anisotropy is due to photons with momentum $k_T < 250$ MeV.

In conclusion, we have shown that model simulations in the 
framework of the linear sigma model predict that the $\pi^0/\pi$ 
ratio shows an enhancement for pions emitted perpendicular to the 
scattering plane when the formation of DCC domains is aided by 
the chiral anomaly. The effect is most pronounced at low, but
non-vanishing transverse pion momenta and at intermediate
heavy ion impact parameters, and for collisions of heavy nuclei.
The anisotropy is also observable in the inclusive photon
spectrum from the decaying neutral pions.

{\em Acknowledgments:}
B.M. wishes to thank the Japan Society for the Promotion of Science 
for support which made his visit to Tokyo Metropolitan University 
possible. H.M. thanks the MIT Center for Theoretical Physics for 
its warm hospitality.  
This work was supported in part by a grant from the
U.~S.~Department of Energy (DE-FG02-96ER90945), and by
Grants-in-Aid for Scientific Research No. 11640271 and
No. 12640285, the Japan Society for the Promotion of Science.


\newpage

\newpage

\begin{references}

\bibitem{DCC}
A.A. Anselm and M.G. Ryskin, {\em Phys. Lett. B{\bf 266}}
(1991) 482;
K.L. Kowalski and C.C. Taylor, preprint hep-ph/9211282.

\bibitem{WA98}
M.M. Aggarwal {\em et al.} (WA98 collaboration),
{\em Phys. Lett. B{\bf 420}} (1998) 169.

\bibitem{Minimax}
T.C. Brooks {\em et al.} (MiniMax collaboration),
{\em Phys. Rev. D{\bf 61}} (2000) 032003.

\bibitem{MM96}
H. Minakata and B. M\"uller,
{\em Phys. Lett. B{\bf 377}} (1996) 135.

\bibitem{AMM98}
M. Asakawa, H. Minakata, and B. M\"uller,
{\em Phys. Rev. D{\bf 58}} (1998) 094011.

\bibitem{FFpi}
A. Anselm, A. Johansen, E. Leader, and L. Lukaszuk, 
{\em Z. Phys. A{\bf 359}} (1997) 457.

\bibitem{Pi96}
R.D. Pisarski,
{\em Phys. Rev. Lett. {\bf 76}} (1996) 3084.

\bibitem{PT97}
R.D. Pisarski and M. Tytgat,
{\em Phys. Rev. D{\bf 56}} (1997) 7077.

\bibitem {KPT98}
D. Kharzeev, R. D. Pisarski, and M. H. G. Tytgat,
Phys. Rev. Lett. {\bf 81} (1998) 512.

\bibitem{olli}
J.-Y. Ollitrault,
{\em Phys. Rev. D{\bf 48}} (1993) 1132.

\bibitem{flow}
N. Herrmann, J.P. Wessels, and T. Wiehold,
{\em Annu. Rev. Nucl. Part. Sci. {\bf 49}} (1999) 581.

\bibitem{sps}
M.M. Aggarwal {\em et al.} (WA98 collaboration),
{\em Phys. Lett. B{\bf 403}} (1997) 390;
{\em Nucl. Phys. A{\bf 638}} (1998) 459c;
H. Appelsh\"auser et al. (NA49 collaboration),
{\em Phys. Rev. Lett. {\bf 80}} (1998) 4138.

\bibitem{star}
K.H. Ackermann (STAR collaboration),
{\em Phys. Rev. Lett. {\bf 86}} (2001) 402.

\end{references}
\end{document}